# Oriented polar molecules trapped in cold helium nanodroplets: Electrostatic deflection, size separation, and charge migration

John W. Niman[1], Benjamin S. Kamerin[1], Daniel J. Merthe[1*], Lorenz Kranabetter[2], and Vitaly V. Kresin[1]

[1]*Department of Physics and Astronomy, University of Southern California,
Los Angeles, CA 90089-0484, USA*

[2]*Institut für Ionenphysik und Angewandte Physik, Universität Innsbruck,
Technikerstr. 25, A-6020 Innsbruck, Austria*

Helium nanodroplets doped with polar molecules are studied by electrostatic deflection. This broadly applicable method allows even polyatomic molecules to attain sub-Kelvin temperatures and nearly full orientation in the field. The resulting intense force from the field gradient strongly deflects even droplets with tens of thousands of atoms, the most massive neutral systems studied by beam "deflectometry." We use the deflections to extract droplet size distributions. Moreover, since each host droplet deflects according to its mass, spatial filtering of the deflected beam translates into size filtering of neutral fragile nanodroplets. As an example, we measure the dopant ionization probability as a function of droplet radius and determine the mean free path for charge hopping through the helium matrix. The technique will enable separation of doped and neat nanodroplets and size-dependent spectroscopic studies.

[*] Present address: Modern Electron, Bellevue, WA 98007, USA

*Introduction.*— If the internal and relative motion of molecules is cooled into the sub-Kelvin range, it becomes possible to observe and steer their reactions with precision, to determine their physical parameters and structures with high accuracy, and to use external fields to finely control their motion and orientation [1-4]. For example, buffer-gas cooling [5] can be employed as an entryway to electrostatic guiding and ultracold trapping [6,7], merged beams enable exploration of chemical reactions in the quantum regime [8], and Stark deflection of small molecules in a supersonic beam can be used to spatially separate their low rotational states and conformers [9].

While high level of control has been demonstrated for individual small molecules, pursuing it for larger polyatomic systems becomes increasingly demanding [10]. Their rotational spectra are more congested, their degrees of freedom are less efficiently and uniformly cooled by nozzle expansion [10,11], and their higher masses reduce the deflection.

A powerful tool to cool and study molecules of a wide range of sizes is "helium nanodroplet isolation" [12-16]. Molecules are entrapped and transported by a beam of $^4$He$_N$ nanodroplets generated by expansion of helium gas through a cryogenic nozzle. Nanodroplets cool by evaporation upon exiting the nozzle, reaching an internal temperature of only $T_0$=370 mK and turning superfluid. This temperature is set by the surface binding energy of helium atoms [17,18] and has been verified, as has the onset of superfluidity, by rotational spectroscopy of entrapped molecules [12]. When the droplet beam passes through one or more vapor-filled cells, atoms and molecules are readily picked up, cooled by heat transfer to the helium matrix (evaporation of surface helium atoms promptly brings the complex back to $T_0$), and carried along by the droplet beam.

This method is unique in being applicable to a variety of molecules and atoms: essentially all that is required for embedding is the availability of ~$10^{-6}$-$10^{-4}$ mbar of vapor. Its other key feature is that it cools all the degrees of freedom of the dopants and ensures that only their lowest vibrational, and in some cases even rotational, levels are occupied. Quantum effects in bimolecular reactions can already become pronounced at $T_0$ (e.g., [19]) and may remain undisrupted by the viscosity-free superfluid matrix. Furthermore, by using sequential pickup it is possible to co-embed multiple (identical or distinct) atoms or molecules in order to explore their interactions and to generate novel or metastable complexes that would be unobtainable by other means.

In the context of control and manipulation by external fields, consider nanodroplet embedding of polar molecules. The salient fact is that by cooling to $T_0$ in this superfluid environment, they become cold enough to strongly (often almost fully) orient themselves along an applied static electric field [20]. Their rotations transform into "pendular" states, employed in landmark spectroscopic studies [34] (see also the recent review [35]). Molecular alignment effects within helium nanodroplets also were recently demonstrated using short laser pulses [36].

Here we subject these systems to the method of electrostatic deflection [9,37-39]. A doped nanodroplet beam passes through an inhomogeneous electric field and its resulting deflection is measured with high accuracy. The attractiveness of such a measurement is that it can be performed using a broad array of molecules (diatomic, polyatomic, complex, agglomerates) and directly yields quantitative observables, without needing to refer to a potentially complex spectroscopic analysis.

However, two potential problems must be considered. First, deflecting neutral droplets by a measurable amount may appear simply unworkable. Indeed, in typical experiments on beams of individual polar molecules or clusters the deflection is at most by a few milliradians, more commonly a fraction of that (translating into millimeters, or fractions thereof, displacement at the detector). Consequently, loading a molecule with a massive coat of barely polarizable helium ought to result in undetectable deflections. Second, the nanodroplets are not identical. Their size distribution is generally log-normal, as is typical of particle growth processes, with a mean $\bar{N}$ that can be shifted by varying the expansion conditions. Can this hinder a deflection experiment?

We report on two principal results. First, we demonstrate that electrostatic deflection of nanodroplets doped with polar molecules is not merely measurable, as we saw in [40], but turns out to be remarkably strong. This is due to the aforementioned orientation effect: when the dipoles' rotational motion is frozen out and they point along the field axis, the resulting great increase in the deflecting force can easily compensate for the additional helium mass. Such a robust effect, in combination with the fact that these may be the most massive (tens of thousands of Daltons) neutral beams subjected to "molecular deflectometry" to date, is noteworthy. The magnitude of the deflections implies that they can be employed for accurate measurements of the dipole moments of complex molecules and to segregation of doped and undoped nanodroplets.

Second, we demonstrate that instead of hindering deflection analysis, droplet size spread can be turned into an informative resource. We show that deflection measurements can be employed to calibrate the nanodroplet size distribution. Even more valuably, deflection can be used to achieve droplet mass filtering, by spatially dispersing the nanodroplets according to their size. This establishes a novel way to perform spectroscopic experiments on neutral nanodroplets as a function of size. As an application and illustration of this method, we study the droplet size dependence of dopant ionization probabilities and determine the mean free path for the migration of positive charge ($He^+$ hole) through the liquid helium matrix.

*Method.*— A supersonic nanodroplet jet is generated by expanding He gas at 80 bar pressure through a cryogenic nozzle, and passes through a pick-up cell positioned downstream. This methodology is described in the cited reviews, see also [40]. Dopants chosen for the present work are dimethyl sulfoxide $(CH_3)_2SO$ ("DMSO," $p=4.0$ Debye) and CsI ($p=11.7$ D). The beam is subsequently collimated by a 0.25 mm × 1.25 mm slit and travels through the 2.5 mm gap between two 15 cm-long electrodes which create an inhomogeneous electric field of the "two-wire" geometry [41-43]. As formulated above, the field orients the polar molecule while its gradient exerts a strong deflecting force on this oriented dipole. The field and gradient strengths range up to ≈85 kV/cm and 350 kV/cm$^2$, respectively.

Approximately 1.3 m past the electrodes the beam enters an electron-impact ionizer (set to 90 eV) through a 0.25 mm-wide slit, and the resultant ions are detected by a quadrupole mass analyzer. The arrival of a doped nanodroplet is registered by setting the analyzer to one of the characteristic fragment peaks of the dopant [44]. In order to isolate the beam-carried signal, the analyzer's output is read via a lock-in amplifier synchronized with a rotating wheel chopper. Additionally, the phase delay between the chopper and analyzer outputs yields the beam velocity $v$ (which rises from 375 m/s at 15 K nozzle temperature to 415 m/s at 19 K). Importantly for deflection measurements, the velocity spread is very narrow, 1-1.5% [45,46]. The deflection angle of a nanodroplet is the ratio of the sideways impulse it receives while traversing the field, $F_z \Delta t \propto \langle p_z \rangle (\partial \mathcal{E}_z / \partial z) v^{-1}$, to its original forward momentum, $mv$. Since the field gradient is proportional to the deflection plate voltage $V$, the droplet's deflection is $d=C\langle p_z\rangle V/(mv^2)$, where $C$ is a constant calculated from the apparatus geometry.

In monitoring the dopant peak in the mass spectrum, one needs to be certain that it is not a fragment of a larger agglomerate deriving from the pick-up of multiple molecules. The probability of embedding $k$ dopants is approximately Poissonian [12]: $P_k = \langle k \rangle^k \exp(-\langle k \rangle)/k!$. Here $\langle k\rangle$ is the average number of pick-up collisions, proportional to the vapor density. Therefore the cell vapor pressure must be low enough for $P_{k>1}$ to remain small. For DMSO we adjust it to produce a usable monomer signal (we use the 78 amu ion peak for deflection measurements) while minimizing the corresponding dimer signal (156 amu). For CsI the procedure is analogous, reinforced by the fact that the Cs$^+$ peak which we use predominantly derives from dissociative ionization of the CsI monomer but not of larger clusters [47].

*Deflections.*— Beam profiles in the detector plane are recorded by measuring the intensity of the chosen ion peak as a function of the ionizer entrance slit position.

Our initiatory deflection measurements scanned this entrance slit in front of the quadrupole's ionizer and suggested beam deflections on the order of a few tenths of a mrad (translating into shifts of a few tenths of a mm in our apparatus) [40]. This was already substantial, but further examination revealed that the actual deflections were considerably larger: we discovered that they extended all the way to the edge of the ionizer's entrance aperture and were artificially clipped there. In order to accommodate such large displacements we now fix the slit in the middle of the aperture and place the entire detector chamber onto a precision linear slide. This enables us to obtain accurate beam profiles extending as far as ±20 mm (16 mrad) from the central axis, see Figs. 1(a,b). These figures show the deflections for a diatomic and a polyatomic dopant, both cooled to 0.37 K by immersion in the superfluid droplet, confirming the broad applicability of the technique.

Such profiles contain a wealth of information. For example, Fig. 1(c) shows that the average deflection is proportional to the deflector voltage. This is fundamentally different from the linear susceptibility regime where $\langle p_z\rangle \propto \mathcal{E}_z$ and therefore $d \propto \mathcal{E}_z \cdot (\partial\mathcal{E}_z/\partial z) \propto V^2$, as commonly observed in cluster beam experiments [37-39]. The dependence plotted here implies saturated susceptibility and provides unambiguous proof that the dopant dipoles are strongly oriented by the applied field [20].

*Nanodroplet sizes and size filtering.*— Each measured profile represents the convolution of single nanodroplet deflections with (a) the distribution of droplet masses, (b) the distribution of their pick-up and ionization cross sections, and (c) the shape of the original undeflected beam. By fitting these profiles to a simulation of the pick-up, deflection, and detection steps [20], we deduce the mean $\bar{N}$ and the width $\Delta N$ of the droplet size distribution produced by the nozzle. As shown in Fig. 1(d), these parameters are in excellent agreement with the standard literature values [12,48]. This both validates our analysis and extends the droplet size calibration curve.

An inspection of Figs. 1(a,b) reveals that the electric field not only shifts the doped droplet beam profile, but also makes it asymmetric. The reason is that smaller, lighter droplets deflect stronger than larger, heavier ones. This immediately suggests that spatial filtering of the deflected beam will translate into size filtering of the neutral nanodroplets.

The ability to scan through nanodroplet sizes within the same beam and within a single experiment is highly appealing, making it possible to explore the influence of droplet size on the spectroscopy and dynamics of embedded molecules. Compared to milestone experiments on droplet sizing by crossed-beam scattering [48,49], here the deflection angles, the size range, and the intensity of the deflected beam are all markedly higher.

*Charge migration.*— To illustrate this capability, we investigate charge migration as a function of droplet size. Consider the steps involved in electron-impact ionization of doped droplets [16,50]. Since helium atoms surround and greatly outnumber the dopant, an electron strike predominantly results in the creation of a $He^+$ ion. The positive hole then resonantly hops from one adjacent helium atom to another, toward the impurity in the middle, until one of two outcomes occurs: it "self-traps" by forming $He_2^+$ followed by the nucleation of larger $He_n^+$ cluster ions, or it reaches and ionizes the dopant. Both outcomes are accompanied by significant energy release which boils off the helium and ejects the ion from the nanodroplet [16].

It follows that the probability of dopant ion formation can be viewed according to Beer's law: $P_m = \exp(-\mathcal{R}/\lambda)$, where $\mathcal{R}$ is the distance which the positive charge needs to travel before reaching the impurity and $\lambda$ is its mean free path before self-trapping. Since $\mathcal{R}$ ~ droplet radius $R$ (see below), a measurement of $P_m$ as a function of droplet size will yield the important physical parameter $\lambda$.

The concept of the measurement is as follows. If a droplet undergoes electric deflection, this automatically implies that it carries an impurity molecule. However, in the mass spectrum it can register either at the impurity mass or at the helium fragment mass. The ratio between these two outcomes, which is precisely $P_m$, can be traced as a function of the droplet's deflection – i.e., of its size.

We developed a procedure to subtract the undoped beam's contribution to the signal and to fit the dopant ion yield to the exponential form $P_m \propto \exp(-\gamma N^{1/3})$ [20]. Fig. 2 assembles the results of measurements using CsI doping performed at three nozzle temperatures, i.e., for strongly distinct $\bar{N}$. Accordingly, it spans a wide range of droplet sizes $N$. It is therefore satisfying that over this full range practically the same value of $\gamma$ (±17%) is found, as anticipated for the ionization pathway described above.

To relate $\gamma$ to the mean free path, we need to estimate $\mathscr{R}$. The He$^+$ hole is originally created at a random location within the droplet [16]. For its subsequent motion, two models can be considered . One [51] assumes that the positive charge hops radially inwards, the other (similar to [52]) that it hops along the dipole's electric field lines all the way from its initial location to the molecule's negative end. Simulating both scenarios and assuming that the dopant occupies a cavity of ≈4 Å radius [53] at the center, we find $\mathscr{R} \approx 0.7R$ for the former case [55] and $\mathscr{R} \approx 1.0R$ for the latter. (This neglects the density gradient near the droplet surface which should not appreciably affect the estimation of the mean free path $\lambda$ [51].) With $R = 2.22 N^{1/3}$ Å [12], this translates into $\mathscr{R} \approx (1.6\text{-}2.2) N^{1/3}$ Å for the two models, respectively, or $\lambda \approx (1.6\text{-}2.2)/\gamma$ Å.

Using the CsI pickup data from Fig. 2 we arrive at $\lambda \approx 16$ Å. (Measurements using DMSO, less accurate and more limited in the spread and assignment of droplet sizes because of its smaller dipole moment and weaker deflections, yielded $\lambda \approx 34$ Å.)

These values are similar to the estimates of 28-35 Å for droplets doped with HCN and HCCCN, found in [51,52] by a very different method: optically selective mass spectrometry. This confirms that our technique is well suited to the task of determining size-dependent parameters of nanodroplet behavior. The referenced estimates are larger than $\lambda$ determined here for CsI, but they were deduced for beams centered at considerably smaller average droplet sizes and containing broad size distributions. This skews the deduced ionization probability, because when the distribution is broad the smaller droplets within it will yield a higher proportion of the impurity ions. This is avoided in the present approach which scans through much narrower nanodroplet size groupings by spreading out the full distribution along the deflection axis.

*Conclusions.*— Cold polar molecules entrapped within superfluid helium nanodroplets can be nearly fully oriented by an external electric field. We showed that this can be exploited in beam deflection experiments. Since the electrostatic deflecting force experienced by an oriented molecular dipole becomes extremely large, we observed that an entire beam of massive nanodroplets, containing up to tens of thousands of He atoms, deflects by impressively large angles.

As demonstrated here, if nanodroplets carry a molecule with a known dipole moment the deflection measurement can be used to calibrate the droplet size distribution in the beam. Conversely, by comparing the deflections of a beam doped with a reference molecule and the same beam doped with another species, one can "read out" the dipole moment of the latter in a model-free approach. Since, as emphasized, nanodroplet embedding is applicable to a broad range of molecules (in particular polyatomic and biological) this introduces a correspondingly broad method of measuring molecular dipole moments. (Note that direct measurements on isolated complex molecules began only recently [37] and many tabulated values still come from liquid phase data with potentially significant uncertainties [56].)

The same approach can be employed with interesting and unusual agglomerates produced via sequential pick-up (this also is a unique capability of helium nanodroplet embedding). For example, it can detect the formation of novel metastable assemblies of cold polar molecules, as we have demonstrated for DMSO dimers and trimers [57]. It should also be usable for the identification of polar vs. nonpolar conformers (cf. [37]).

Speaking of molecular characterization, it should also be possible to use strong electrostatic deflections to separate doped and undoped nanodroplets, which is important for emerging applications aiming at structural analysis of embedded molecules by x-ray, EUV, and electron pulses [15,58-60].

Finally, we pointed out that since the deflection angle of a doped nanodroplet depends on its mass, the broad size distribution contained within the original beam becomes spatially spread out by the time it reaches the detector plane. In other words, the deflection process disperses the He$_N$ population and establishes a means to probe the behavior of neutral nanodroplets as a function of their size *N*. To illustrate this, we measured how the dopant ion yield varies with droplet radius, and thus determined the mean free path for the migration of positive charge through the helium matrix. This droplet size-filtering technique will make possible size-resolved spectroscopy of cold dopants and dopant reactions.

*Acknowledgements.* This work was supported by the U. S. National Science Foundation under Grant No. CHE-1664601. L.K. would like to acknowledge a scholarship from the Austrian Marshall Plan Foundation and support from the Austrian Science Fund under project FWF W1259. We thank Jiahao Liang and the staff of the USC Machine Shop for their help with the project.

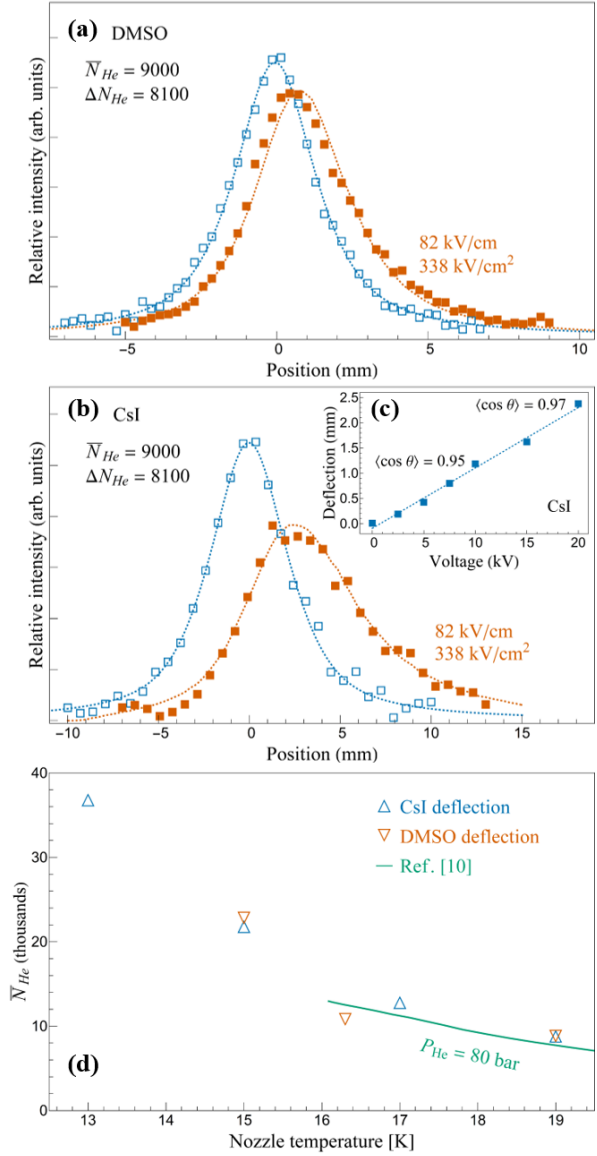

**FIG. 1**. (a) Deflection of He$_N$ nanodroplets with DMSO dopant. Squares: experimental data; blue line: pseudo-Voigt function fit to the undeflected profile; red line: fit by simulation of the deflection process [20]. (b) Same for CsI dopant. (c) Average deflection of the nanodroplets beam vs. electrode voltage. Its linear variation attests to the strong orientation of the cold dopant molecule along the field, cf. calculated orientation cosine labels. (d) Average nanodroplet size as a function of nozzle temperature. Symbols: mean, $\bar{N}$, of the log-normal size distribution deduced from our deflection measurements; line: data from Ref. [12]. Fit to the deflection data yielded $\Delta N/\bar{N} = 0.85$ for the FWHM of the distribution, in excellent agreement with Ref. [48].

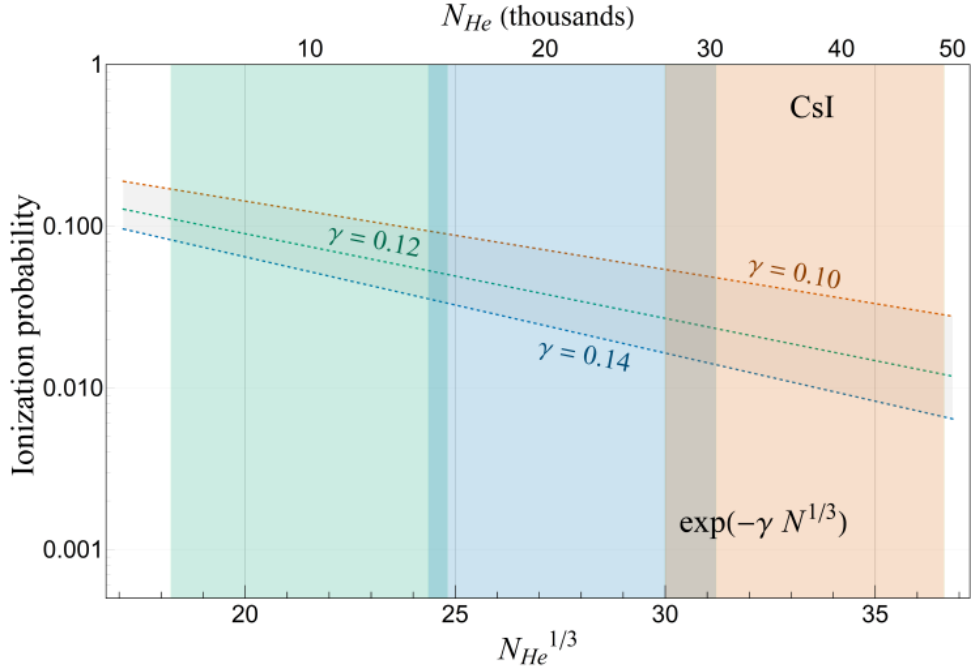

**FIG. 2**. Probability of ionizing charge transfer to embedded CsI molecules as a function of nanodroplet size. This probability was determined by a fit to the electric deflection profiles which spatially disperse nanodroplets according to their mass, as described in the text. The displayed size range was covered by measurements at three nozzle temperatures (13 K, 15 K, and 19 K) corresponding to average droplet sizes $\bar{N}$ of $3.7\times10^4$, $2.2\times10^4$ and $9\times10^3$ atoms, respectively [see Fig. 1(d)]. At each of these temperatures the beam contained a log-normal distribution of droplet sizes which then spread out spatially upon deflection. This allowed the data to span a range of sizes, as marked by the three (overlapping) bands of color. For each of those bands the probability of dopant ion formation was fitted to the form $P_m \propto \exp(-\gamma N^{1/3})$. The results are depicted as dashed lines in the figure, color-matched to the size band from which the corresponding value of $\gamma$ was derived. The lines extend into neighboring bands in order to show the range of uncertainty in their slope; the fact that they are close demonstrates the consistency of the analysis.

REFERENCES

[1] *Low Temperatures and Cold Molecules*, ed. by I. W. M. Smith (Imperial College Press, London, 2008).

[2] *Cold Molecules: Theory, Experiment, Applications*, ed. by R. V. Krems, W. C. Stwalley, and B. Friedrich (CRC Press, Boca Raton, 2009).

[3] *Cold Chemistry: Molecular Scattering and Reactivity Near Absolute Zero*, ed. by O. Dulieu and A. Osterwalder (Royal Society of Chemistry, Cambridge, 2018).

[4] M. Lemeshko, R. V. Krems, J. M. Doyle, and S. Kais, Mol. Phys. **111**, 1648 (2013).

[5] N. R. Hutzler, H.-I. Lu, and J. M. Doyle, Chem. Rev. 112, 4803 (2012).

[6] L. D. van Buuren, C. Sommer, M. Motsch, S. Pohle, M. Schenk, J. Bayerl, P. W. H. Pinkse, and G. Rempe, Phys. Rev. Lett. **102**, 033001 (2009).

[7] J. L. Bohn, A. M. Rey, and J. Ye, Science **357**, 1002 (2017).

[8] E. Lavert-Ofir, Y. Shagam, A. B. Henson, S. Gersten, J. Kłos, P. S. Żuchowski, J. Narevicius, and E. Narevicius, Nat. Chem. **6**, 332 (2014).

[9] Y.-P. Chang, D. Horke, S. Trippel, and J. Küpper, Intern. Rev. Phys. Chem. **34**, 557 (2015).

[10] F. Filsinger, J. Küpper, G. Meijer, L. Holmegaard, J. H. Nielsen, I. Nevo, J. L. Hansen, and H. Stapelfeldt, J. Chem. Phys. **131**, 064309 (2009).

[11] B. K. Stuhl, M. T. Hummon, and J. Ye, Annu. Rev. Phys. Chem. **65**, 501 (2014).

[12] J. P. Toennies and A. F. Vilesov, Angew. Chem. Int. Ed. **43**, 2622 (2004).

[13] F. Stienkemeier and K. K. Lehmann, J. Phys. B: At., Mol. Opt. Phys. 39, R127 (2006).

[14] S. Yang and A. M. Ellis, Chem. Soc. Rev. 42, 472 (2013).

[15] R. M. Tanyag, C. F. Jones, C. Bernando, S. M. O. O'Connell, D. Verma, and A. F. Vilesov, in Ref. [3], p. 389.

[16] A. Mauracher, O. Echt, A. M. Ellis, S. Yang, D. K. Bohme, J. Postler, A. Kaiser, S. Denifl, and P. Scheier, Phys Rep. **751**, 1 (2018).

[17] D. M. Brink and S. Stringari, Z. Phys. D **15**, 257 (1990).

[18] K. Hansen, *Statistical Physics of Nanoparticles in the Gas Phase*, 2nd ed. (Springer, Cham, 2018).

[19] T. V. Tscherbul and R. V. Krems, J. Chem. Phys. **129**, 034112 (2008).

[20] See the Supplemental Material, which includes Refs. [21-33], for details on calculating the molecular orientation, fitting the measured deflection profiles, and analyzing the charge transfer probability data.

[21] J. E. Wollrab, *Rotational Spectra and Molecular Structure* (Academic, New York, 1967).

[22] R. N. Zare, *Angular Momentum* (Wiley, New York, 1988).

[23] Y.-P. Chang, F. Filsinger, B. G. Sartakov, and J. Küpper, Comput. Phys. Commun. **185**, 339 (2014).

[24] H. Dreizler and G. Dendl, Z. Naturforsch. A **19**, 512 (1964).

[25] M. L. Senent, S. Dalbouha, A. Cuisset, and D. Sadovskii, J. Phys. Chem. A **119**, 9644 (2015).

[26] B. Friedrich and D. Herschbach, Int. Rev. Phys. Chem. **15**, 325 (1996).

[27] J. Bulthuis, J. A. Becker, R. Moro, and V. V. Kresin, J. Chem. Phys. **129**, 024101 (2008).

[28] M. H. G. de Miranda, A. Chotia, B. Neyenhuis, D. Wang, G. Quemener, S. Ospelkaus, J. L. Bohn, J. Ye, and D. S. Jin, Nat. Phys. **7**, 502 (2011).

[29] M. Lemeshko, Phys. Rev. Lett. **118**, 095301 (2017).

[30] M. Yu. Skripkin, P. Lindqvist-Reis, A. Abbasi, J. Mink, I. Persson, and M. Sandström, Dalton T. **23**, 4038 (2004).

[31] A. L. Stancik and E. B. Brauns, Vib. Spectrosc. **47**, 66 (2008).

[32] E. H. P. Cordfunke, Thermochim. Acta **108**, 45 (1986).

[33] B. E. Callicoatt, K. Förde, L. F. Jung, T. Ruchti, and K. C. Janda, J. Chem. Phys. **109**, 10195 (1998).

[34] M. Y. Choi, G. E. Douberly, T. M. Falconer, W. K. Lewis, C. M. Lindsay, J. M. Merritt, P. L. Stiles, and R. E. Miller, Int. Rev. Phys. Chem. **25**, 15 (2006).

[35] D. Verma, R. M. P. Tanyag, S. M. O'Connell, and A. F. Vilesov, Adv. Phys.: X **4**, 1553569 (2019).

[36] A. S. Chatterley, C. Schouder, L. Christiansen, B. Shepperson, M. Heidemann Rasmussen, and H. Stapelfeldt, Nat. Commun. **10**, 133 (2019).

[37] M. Broyer, R. Antoine, I. Compagnon, D. Rayane, and Ph. Dugourd, Phys. Scr. **76**, C135 (2007).

[38] W. A. de Heer and V. V. Kresin, in *Handbook of Nanophysics: Clusters and Fullerenes*, ed. by K. D. Sattler (CRC Press, Boca Raton, 2010).

[39] S. Heiles and R. Schäfer, *Dielectric Properties of Isolated Clusters:Beam Deflection Studies* (Springer, Dordrecht, 2014).

[40] D. J. Merthe and V. V. Kresin, J. Phys. Chem. Lett. **7**, 4879 (2016).

[41] N. F. Ramsey, *Molecular Beams* (Oxford University Press: Oxford; 1956).

[42] G. Tikhonov, K. Wong, V. Kasperovich, and V. V. Kresin, Rev. Sci. Instrum. **73**, 1204 (2002).

[43] D. J. Merthe, PhD dissertation, University of Southern California, 2017.

[44] *NIST Chemistry WebBook*, NIST Standard Reference Database No. 69, ed. by P. J. Linstrom and W. G. Mallard (National Institute of Standards and Technology, Gaithersburg), http://webbook.nist.gov.

[45] H. Buchenau, E. L. Knuth, J. Northby, J. P. Toennies, and C. Winkler, J. Chem. Phys. **92**, 6875 (1990).

[46] J. Harms, J. P. Toennies, and E. L. Knuth, J. Chem. Phys. **106**, 3348 (1997).

[47] M. F. Butman, L. S. Kudin, A. A. Smirnov, and Z. A. Munir, Int. J. Mass Spectrom. **202**, 121 (2000).

[48] J. Harms, J. P. Toennies, and F. Dalfovo, Phys. Rev. B **58**, 3341 (1998).

[49] M. Lewerenz, B. Schilling, and J. P. Toennies, Chem. Phys. Lett. **206**, 381 (1993).

[50] A. Scheidemann, B. Schilling, and J. P. Toennies, J. Phys. Chem. **97**, 2128 (1993).

[51] A. M. Ellis and S. Yang, Phys. Rev. A **76**, 032714 (2007).

[52] W. K. Lewis, C. M. Lindsay, R. J. Bemish, and R. E. Miller, J. Am. Chem. Soc. **127**, 7235 (2005).

[53] Estimate based on ionic radii and on optical data [54].

[54] P. Pacak, J. Solution Chem. **16**, 71 (1987).

[55] This coefficient has a straightforward origin: for charges created at a random position within a sphere, the average distance from the center (neglecting the small dopant cavity) is $V^{-1}\int_0^R r\,dV = 0.75R$.

[56] D. R. Lide, "Dipole Moments," in *CRC Handbook of Chemistry and Physics*, 99th ed., ed. by J. R. Rumble (CRC Press, Boca Raton, 2018).

[57] J. W. Niman, B. S. Kamerin, L. Kranabetter, D. J. Merthe, J. Suchan, P. Slavíček, and V. V. Kresin (unpublished).

[58] L. F. Gomez *et al.,* Science **345**, 906 (2014).

[59] D. Rupp *et al*., Nat. Commun. **8**, 493 (2017).

[60] Y. He, J. Zhang, and W. Kong, J. Chem. Phys. **145**, 034307 (2016).

# Oriented polar molecules trapped in cold helium nanodroplets: Electrostatic deflection, size separation, and charge migration

J. W. Niman[1], B. S. Kamerin[1], D. J. Merthe[1], L. Kranabetter[2], and V. V. Kresin[1]

[1]*Department of Physics and Astronomy, University of Southern California, Los Angeles, CA 90089-0484, USA*

[2]*Institut für Ionenphysik und Angewandte Physik, Universität Innsbruck, Technikerstr. 25, A-6020 Innsbruck, Austria*

## I. MOLECULAR ORIENTATION

The orientation of a rigid polar molecule in an external electric field derives from the rotational Stark effect, a topic extensively described in the literature (see, e.g., [S1,S2]). This section provides an overview of several points which are relevant to the current experiment. The starting Hamiltonian for a rotor with an electric dipole moment $p$ in an electric field $\vec{\mathcal{E}} = \mathcal{E}\hat{z}$ is

$$\hat{H} = A\hat{J}_a^2 + B\hat{J}_b^2 + C\hat{J}_c^2 - p\mathcal{E}\cos\theta, \tag{S.1}$$

where $A$, $B$, $C$ are the rotational constants (reciprocals of the moments of inertia) for the molecule's principal axes $a,b,c$, and $\hat{J}$ are the angular momentum components corresponding to the same axes. The angle between the dipole moment and the field is denoted by $\theta$; the Stark term also can be rewritten in terms of the dipole moment and electric field components in the body-fixed frame of the molecule. The most convenient basis for diagonalizing this Hamiltonian is the set of Wigner $D$-matrices. The corresponding matrix elements are listed, e.g., in [S3].

The orientation cosine of a molecule in an energy eigenstate with energy $E(\mathcal{E})$ is given by

$$\langle\cos\theta\rangle = \frac{\langle p_z\rangle}{p} = -\frac{1}{p}\frac{\partial E}{\partial \mathcal{E}}, \qquad (S.2),$$

and the thermally averaged orientation at a rotational temperature $T$ is

$$\langle\cos\theta\rangle = \frac{k_B T}{p}\frac{\partial}{\partial \mathcal{E}}\left(\ln\sum_E e^{-E(\mathcal{E})/k_B T}\right) \qquad (S.3)$$

Fig. S1 shows the Stark level diagram for molecules studied in this work (the rotational constants are 0.235 cm$^{-1}$, 0.231 cm$^{-1}$, and 0.141 cm$^{-1}$ for DMSO [S4,S5], and 0.0236 cm$^{-1}$ for CsI [S6]) and their orientation at the nanodroplet temperature of 0.37 K. At high electric field strengths the orientation cosine approaches unity, which means that the molecule oscillates around the electric field direction, corresponding to a pendular state.

The right-hand panels of Fig. S.1 also include the classical Langevin function

$$\langle\cos\theta\rangle = \coth x - 1/x, \qquad (S.4)$$

where $x \equiv p\mathcal{E}/k_B T$. This becomes a good approximation when $T$ is much higher than the rotational constants [S7,S8], meaning that many rotational states are occupied. As the figure demonstrates, this is indeed the case for the heavy rotor CsI whose rotational constant is equivalent to 0.034 K in temperature units, and remains a passable approximation even for DMSO. This means that the measurements described in this letter detect the average deflection of a superposition of rotational states, as opposed to some of the cold-molecule studies cited in the main text which focus on resolving individual quantum state deflections. (We emphasize, though, that we use quantum-mechanical diagonalization and not the classical function in modeling the nanodroplet deflection process as described in Section S.II below.)

We now can elaborate on the plot of average beam deflection vs. electrode voltage, Fig. 1(c) of the main text. For rotational temperatures above several K and practical electric field strengths, rotational motion severely suppresses the polarizing action of the external field: the ratio $x$ remains small even for dipole moments of several Debye. To illustrate this point we can refer to the Langevin function: $\langle p_z \rangle \xrightarrow[x \ll 1]{} px/3 = p \cdot [p\mathcal{E}/(3k_B T)]$. In this case, as stated in the main text, experiments find the average deflection to be proportional to the square of the applied deflection voltage (one power for $\mathcal{E}$ in the above formula and one power for its gradient, i.e., the deflecting force). It is only when $T$ becomes very low, as enabled here by helium nanodroplet isolation, that the orientation can approach saturation: $\langle p_z \rangle \to p$ and the deflection now varies only with the strength of the field gradient, i.e., linearly with the applied voltage $V$. This is precisely what is confirmed by Fig. 1(c) in juxtaposition with Fig. S1(c): above an applied voltage of 2.5 kV (corresponding to $\mathcal{E}$=10 kV/cm) the deflection of CsI doped nanodroplets follows a straight line. [The minor nonlinearity at lower fields in Fig. 1(c) reflects the remaining small rise of the orientation curve in Fig. S1(c).]

Of course, if molecules can be made even colder then even weaker dipoles can be oriented with even weaker fields. For example, full orientation of KRb ($p$=0.16 D) in an optical trap at temperatures below 1 μK was achieved with a field of 4 kV/cm [S9]. But a crucial point about superfluid nanodroplets is that by offering a "universal" trap at $T_0$=0.37 K they make it possible to strongly orient those molecules which are too large and complex for optical trapping methods.

So far in the discussion the role of the matrix has been limited to that of a thermal bath. Although superfluid helium is nondissipative, its interaction with dopant molecules has been found to increase their effective rotational constants [S10-S12]. This effect is mostly observed for anisotropic molecules with rotational constants $\lesssim 1$ cm$^{-1}$ and on average decreases them by a factor of ~2.5. This increases the ratio of the temperature to the rotational constant and the molecule effectively behaves more "classically," acquiring a stronger orientation in the electric field. At the field strength used here, Figs. S1(c,d) show that the orientation cosine of CsI will be barely affected by such a shift, while that of DMSO may move somewhere between the two lines (no experimental or theoretical values are available for the renormalization of its rotational constant) for an approximate resulting uncertainty in $\langle\cos\theta\rangle$ of ±10%.

A quantum phenomenon predicted for superfluid droplets is the appearance of "pendulons" (librational states dressed by a field of many-particle excitations) with a spectrum exhibiting a series of instabilities tunable by the applied electric field. An interesting question is whether they may have an observable influence on field deflection profiles.

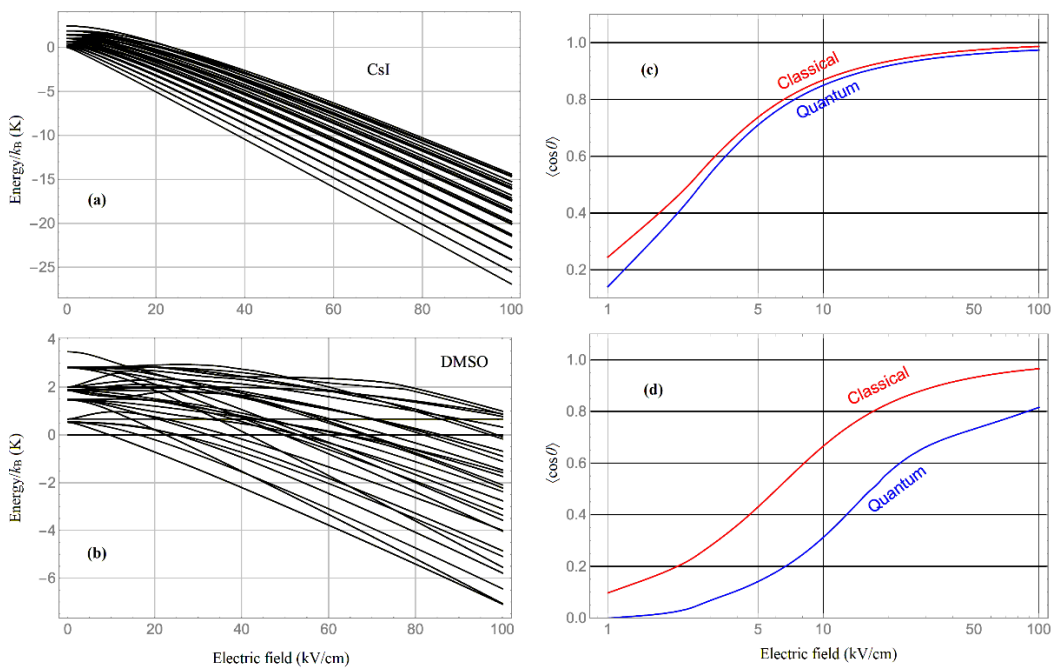

**FIG. S1**. (a,b) Rotational Stark levels and (c,d) alignment of CsI and DMSO molecules at $T$=0.37 K. The blue and red curves show the thermally averaged orientation of the quantum and classical rotors. Deflections used in the present work were primarily acquired at 82 kV/cm.

## II. DEFLECTION PROFILES

For a quantitative analysis of doped nanodroplet deflection a Monte Carlo simulation of the process is implemented. An earlier version was described in [S13,S14], and since then has undergone substantial further development. The simulation starts by picking a $^4\text{He}_N$ nanodroplet from a log-normal distribution of sizes and assigning it an initial longitudinal velocity $v$ based on the beam velocity measurement. The droplet's capture of a dopant molecule results in the deposition of a certain amount of energy which is composed of the following ingredients: (*i*) the kinetic energy lost in the inelastic collision between the dopant and the droplet (averaged over all relative collision angles and molecular speeds), (*ii*) the molecule's thermally averaged rotational energy, and (*iii*) the vibrational energy contained in the molecule's normal modes [S15,S6]. This energy is dissipated by the evaporation of He atoms from the droplets (0.5-0.6 meV energy release per He atom [S10]): on the order of a thousand atoms are boiled off. The probability of a pick-up collision is weighed according to the droplet's geometric cross section (i.e., with an $N^{2/3}$ weight).

The doped nanodroplet, cooled by prompt evaporation back to 0.37 K, then enters the electric field region. The rigid-rotor Stark effect Hamiltonian of the dopant is diagonalized numerically [S13] (cf. [S3]) and the orientation of its dipole moment, $\langle p_z \rangle$, is found by sampling from the in-field Boltzmann distribution at the droplet temperature. A small correction for the electric polarization of the helium matrix is added to the effective dipole moment [S14]. The electric plates' calculated field gradient [S16] is used to compute the deflection force on the droplet during its passage through the field. The droplet is then allowed to drift through the free flight path until it arrives at the detector entrance. This procedure is repeated multiple times to generate approximately half a million detection events, simulating the shape distribution of the beam flux arriving at the detector.

The calculated distribution is convolved with the beam profile measured at zero electric field. In this way we account for the initial small transverse velocity of the nozzle beam and for the transmission function of the ionizer entrance slit. The output is then weighed according to the droplets' electron impact ionization cross section ($\propto N^{2/3}$) and the molecular ion formation probability $P_m \propto \exp(-\gamma N^{1/3})$. The latter function is described in the main text and the determination of $\gamma$ is described in Section III of this Supplemental Material. The result can now be fitted to the experimentally measured deflection profile of the polar dopant's ion signal. (More precisely, the experimental deflection profile is first represented by a Pseudo-Voigt function with a sigmoidal width parameter [S17] which matches the shape of the data very well and mitigates scatter in the intensity counts, especially near the profile edges.) As explained at the end of Section III, the fitting procedure is run iteratively, the fitting parameters being the mean droplet size $\bar{N}$ in the log-normal distribution of sizes, and the formation exponent $\gamma$. The full width at half maximum (FWHM) of the size distribution is set to $\Delta N = 0.85 \bar{N}$ throughout, which was found to provide the best fit to the data and is in excellent agreement with the values reported in Ref. [S18].

## III. Charge transfer probabilities

As emphasized in the main text, a benefit of the deflection method is that if a droplet is identified as having been deflected by the electric field, then – regardless of what ion it produces after electron-impact ionization – one can be sure that it originally contained a polar dopant. In addition, the amount of the droplet's deflection is inversely proportional to its mass. From this one can determine the relative probabilities of the ionization channels of a doped nanodroplet as a function of its size.

However, since the deflected (doped) and undeflected (undoped) beams both have finite widths and spatially overlap, a procedure needs to be formulated to separate the total measured profile intensities into their individual contributions.

Let us turn on the deflecting field and denote the position of the detector entrance collimator (i.e., the profile coordinate) by $x$, the dopant molecule ion signal profile by $S^m(x)$, and the helium ion signal profile by $S^{He}(x)$. The latter can be measured under two conditions: when the pick-up cell is empty and when it is filled with dopant vapor. We label the corresponding profiles as $S_e^{He}(x)$ and $S_v^{He}(x)$.

The helium peaks in the mass spectrum can derive from three sources: (*i*) pure undoped helium droplets; (*ii*) droplets doped with the polar molecule of interest; and also (*iii*) droplets containing a background gas molecule (water, oxygen, nitrogen) unpreventably picked up along the path from the skimmer to the deflection plates. This flight path region is kept at background vacuum levels below $2\times10^{-7}$ torr and the interior of the pick-up chamber is additionally desorbed of water vapor by UV irradiation (UVB-100 system, RBD Instruments, Bend, Oregon). Therefore all the pick-up probabilities are small and we make the assumption that a nanodroplet can contain at most a single impurity, either the polar dopant or a background molecule.

We denote the incoming neutral nanodroplet fluxes corresponding to cases (*i*), (*ii*), and (*iii*) by $I^u(x)$, $I^m(x)$, and $I^b(x)$, respectively. Then the ion signal profiles defined above can be expressed as follows:

$$S_e^{He}(x) = \varepsilon_{He}\left\{I^u(x) + I^b(x)P_{He}^b\right\}, \qquad (S.5)$$

$$S_v^{He}(x) = \varepsilon_{He}\left\{\eta\left[I^u(x) + I^b(x)P_{He}^b\right] + I^m(x)P_{He}^m\right\}, \qquad (S.6)$$

$$S^m(x) = \varepsilon_m I^m(x) P_m^m. \qquad (S.7)$$

Here $P_m^m$ and $P_{He}^m$ respectively represent the probabilities that electron-impact ionization of a nanodroplet containing a polar molecule will produce either a molecular ion or a helium ion. Similarly, $P_{He}^b$ is the probability that a nanodroplet containing a background impurity will yield a helium ion. In principle, all three are functions of nanodroplet radius, however we can make the simplifying assumption that $P_{He}^m \approx P_{He}^b \approx 1$, because for the droplet sizes studied here the self-trapping of the positive charge is a significantly more likely outcome than its reaching and ionizing the dopant. This is consistent with the end result shown in Fig. 2 of the main text. Thus we retain only the droplet size dependence of the latter outcome, which is therefore dependent on the deflection coordinate $P_m^m(x) = P_m^m[N(x)]$.

The coefficients $\varepsilon_{He}$ and $\varepsilon_m$ are the mass spectrometer's detection efficiencies of the helium ions and molecular ions. Since these masses aren't radically different, it is adequate to approximate the efficiencies as equal to each other.

Finally, the factor $\eta$ represents the decreased flux of nanodroplets which do not contain a polar molecule. This decrease derives from various scattering processes of the primary beam in the vapor cell, all of which are proportional to the droplet cross section. Therefore we write $\eta \approx 1 - c\bar{N}^{2/3}$. Physically, the second term should be on the order of $nL\sigma$, where $n$ is the number density of molecular vapor in the pick-up cell [$n=P_{vapor}/(k_B T_{cell})$], $L$ is the path length through the cell, and $\sigma \sim \pi R^2 \approx \pi (2.2 \text{Å})^2 \bar{N}^{2/3}$. With the experimental parameters for CsI ($T$=680 K, $P_{vapor} \approx 5\times 10^{-3}$ Pa [S19], $L \approx 2.75$ cm) we expect $c \approx 10^{-3}$, and this indeed is the average value of $c$ which comes out of the data fits described in the next paragraph. This confirms the applicability of the expression for $\eta$.

With the above approximations, Eqs. (S.5)-(S.7) result in

$$S_v^{He}(x) \approx \left(1 - c\bar{N}^{2/3}\right) S_e^{He}(x) + \frac{S^m(x)}{P_m^m(x)} \tag{S.8}$$

Thus the experimental procedure consists of measuring three beam profiles in the presence of the deflecting electric field: the helium ions' when the pick-up cell is empty, $S_e^{He}(x)$; the helium ions' when the pick-up cell is filled with molecular vapor, $S_v^{He}(x)$; and the picked-up molecules' deflection profile, $S^m(x)$. These are illustrated in Fig. S2. The nanodroplet sizes in the beam and along the deflection profiles being known from the analysis described in Section II above, a fit to Eq. (S.8) yields the desired probability of dopant molecular ion formation $P_m^m[N(x)]$. The main text refers to this function simply as $P_m$ and discusses its $\propto \exp(-\gamma N^{1/3})$ behavior.

Two additional notes are in order. First of all, we have been referring to the "helium ion signal" and to the "dopant molecule ion signal." In actuality the helium droplets produce a mass spectrum consisting of a sequence of $He_n^+$ peaks whose intensity decreases with $n$. Likewise, a polyatomic molecule will produce a pattern of fragment ions. For the fit using Eq. (S.8), we measured the profiles of the most intense peaks ($He_2^+$ and $DMSO^+$) and scaled them by the ratio of these peaks' area to the area under all the fragment ion peaks in the full beam's mass spectrum (≈70% and ≈30%, respectively), cf., e.g., Ref. [S20].

Secondly, the fitting of the deflection profile (Section II) and of the size-dependent ion formation efficiency (Section III) must be performed self-consistently. Consequently, the analysis described in the two sections is performed iteratively and the fits typically converge within a few iterations.

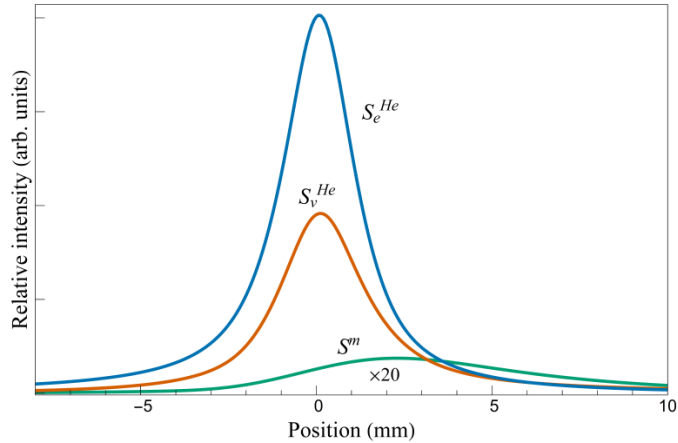

**FIG. S2**. Examples of beam profiles which serve as input to fitting Eq. (S.8) for determining the ionization probability of a molecule embedded in a helium nanodroplet. The curves shown are skewed Pseudo-Voigt envelopes of the experimental beam profiles which were obtained using CsI dopant molecules with nanodroplet source temperature of 19 K and deflection electrode voltage of 20 kV. The curves are labelled according to the variable names used in Eq. (S.8).

## References

[S1]  J. E. Wollrab, *Rotational Spectra and Molecular Structure* (Academic, New York, 1967).

[S2]  R. N. Zare, *Angular Momentum* (Wiley, New York, 1988).

[S3]  Y.-P. Chang, F. Filsinger, B. G. Sartakov, and J. Küpper, Comput. Phys. Commun. **185**, 339 (2014).

[S4]  W. Feder, H. Dreizler, and H. D. Rudolph, Z. Naturforsch. A **24**, 266 (1969).

[S5]  M. L. Senent, S. Dalbouha, A. Cuisset, and D. Sadovskii, J. Phys. Chem. A **119**, 9644 (2015).

[S6]  *NIST Chemistry WebBook*, NIST Standard Reference Database No. 69, ed. by P. J. Linstrom and W. G. Mallard (National Institute of Standards and Technology, Gaithersburg), http://webbook.nist.gov.

[S7]  B. Friedrich and D. Herschbach, Int. Rev. Phys. Chem. **15**, 325 (1996).

[S8]  J. Bulthuis, J. A. Becker, R. Moro, and V. V. Kresin, J. Chem. Phys. **129**, 024101 (2008).

[S9]  M. H. G. de Miranda, A. Chotia, B. Neyenhuis, D. Wang, G. Quemener, S. Ospelkaus, J. L. Bohn, J. Ye, and D. S. Jin, Nat. Phys. **7**, 502 (2011).

[S10] J. P. Toennies and A. F. Vilesov, Angew. Chem. Int. Ed. **43**, 2622 (2004).

[S11] M. Y. Choi, G. E. Douberly, T. M. Falconer, W. K. Lewis, C. M. Lindsay, J. M. Merritt, P. L. Stiles, and R. E. Miller, Int. Rev. Phys. Chem. **25**, 15 (2006).

[S12] M. Lemeshko, Phys. Rev. Lett. **118**, 095301 (2017).

[S13] D. J. Merthe, PhD dissertation, University of Southern California, 2017.

[S14] D. J. Merthe and V. V. Kresin, J. Phys. Chem. Lett. **7**, 4879 (2016).

[S15] M. Yu. Skripkin, P. Lindqvist-Reis, A. Abbasi, J. Mink, I. Persson, and M. Sandström, Dalton T. **23**, 4038 (2004).

[S16] G. Tikhonov, K. Wong, V. Kasperovich, and V. V. Kresin, Rev. Sci. Instrum. **73**, 1204 (2002).

[S17] A. L. Stancik and E. B. Brauns, Vib. Spectrosc. **47**, 66 (2008).

[S18] J. Harms, J. P. Toennies, and F. Dalfovo, Phys. Rev. B **58**, 3341 (1998).

[S19] E. H. P. Cordfunke, Thermochim. Acta **108**, 45 (1986).

[S20] B. E. Callicoatt, K. Förde, L. F. Jung, T. Ruchti, and K. C. Janda, J. Chem. Phys. **109**, 10195 (1998).